# Direct 3D Tomographic Reconstruction and Phase-Retrieval of Far-Field Coherent Diffraction Patterns

T. Ramos[1], Bastian E. Grønager[2], Martin Skovgaard Andersen[3], J. W. Andreasen[2,*]

[1]Technical University of Denmark, Department of Energy Conversion and Storage, Roskilde, Denmark
[2]Technical University of Denmark, Department of Physics, Lyngby Denmark
[3]Technical University of Denmark, Department of Applied Mathematics and Computer Science, Lyngby, Denmark
[*]Corresponding author: jewa@dtu.dk

**Abstract**
**We present an alternative numerical reconstruction algorithm for direct tomographic reconstruction of a sample's refractive indices from the measured intensities of its far-field coherent diffraction patterns. We formulate the well-known phase-retrieval problem in ptychography in a tomographic framework which allows for simultaneous reconstruction of the illumination function and the sample's refractive indices in three dimensions. Our iterative reconstruction algorithm is based on the Levenberg-Marquardt algorithm and we demonstrate the performance of our proposed method with simulated and real datasets.**

## Introduction

The performance and behaviour of numerous engineering materials and biological systems in different scientific fields is largely determined by their internal structure down to the micro and nanoscale. The ability to image such nanoscale structures is considered to be a crucial tool for their study and understanding in order to promote their further development and research. X-ray microscopy is a non-invasive technique well suited for imaging such systems due to its minimal sample preparation requirements, high spatial resolution, high degree of penetration (compared with electron microscopy) possibility for quantitative measurements and *in situ* or *operando* experimental setups. Of the different X-ray microscopy variants, coherent diffraction imaging (CDI) techniques do not rely on X-ray imaging optics such as lenses and their spatial resolution is consequently not limited by lens aberrations or numerical apertures and has the potential to reach atomic length scales comparable to the illumination wavelength. Moreover, by relying on the diffractive properties of coherent radiation, CDI techniques are able to image samples with sizes smaller than the X-ray detector pixel itself and return quantitative information of the sample's complex refractive index which is inaccessible by conventional X-ray transmission methods.

Forward scattering X-ray ptychography, first demonstrated in 2007 [1], is a scanning variant of CDI that aims to relax the requirements for finite sample sizes defined within a compact support, and plane incoming wave front, allowing for increased fields-of-view and the additional reconstruction of the illumination function. In a CDI experiment the incoming beam, described by the probe function, interacts with the sample and is attenuated and phase shifted before being propagated into an X-ray photon counting detector that measures its intensity. Without an imaging lens, any phase information of the wave-field is lost, giving rise to the so called missing-phase problem.

Ptychography or CDI phase retrieval algorithms replace the purpose of an image forming lens by recovering the unknown phase numerically, using iterative algorithms mostly based on some type of optimization scheme. A few exceptions to these methods exist with closed-form solutions [2], [3] but their high sampling requirements make their application practically infeasible.

3D ptychography or ptycho-tomography extends the unique capabilities of CDI techniques to a higher spatial dimension, relying on tomographic reconstruction algorithms to assemble three-dimensional volumes representing the sample's refractive indices, from which the absorption properties and electron density may be deduced. Here, 3D ptychography is to be understood as described and should not be mistaken for the multi-slice reconstruction approach described in [4]. The combination of both CDI and tomographic methods is not direct, involving intermediate data analysis steps mostly concerning phase-unwrapping, background data normalization and tomographic alignment operations. Several methods to address these issues have already been presented in the literature [5]–[9] and have so far been successfully applied in different fields of applications.

As X-ray ptycho-tomography expands into new scientific areas and *in situ* and *operando* studies, the need for faster acquisition times is becoming a decisive factor for the success of an experiment. Besides, new generation synchrotrons will deliver higher fluxes with a superior degree of coherence, and thus the spatial resolution of CDI techniques will most likely be limited by radiation damage suffered by the sample, rather than the coherence properties of the beam. According to the dose fractionation theorem, introduced by Hegerl and Hoppe [10], [11], the dose required to achieve a given statistical significance for each voxel of the three-dimensional tomogram is the same as required to measure a single projection of the same voxel with the same statistical significance. So far, ptychographic reconstruction algorithms rely on a significant overlap between illuminations in different scanning coordinates in order to constrain the domain of possible solutions and accurately recover the phase contrast image. A direct combination of ptychographic reconstruction methods with tomographic algorithms is thought to be able to relax this illumination overlap requirement by reducing its domain to a single 3D volume instead of a series of 2D projections (see Figure 1). Additionally, 2D ptychography images may contain non-negligible reconstruction errors that result in artefacts or resolution deterioration during tomographic reconstruction [12], [13]. Such uncertainties are expected to diminish in a direct ptycho-tomography reconstruction algorithm, because all measured data is implicitly *forced* to be consistent within a single three-dimensional volume.

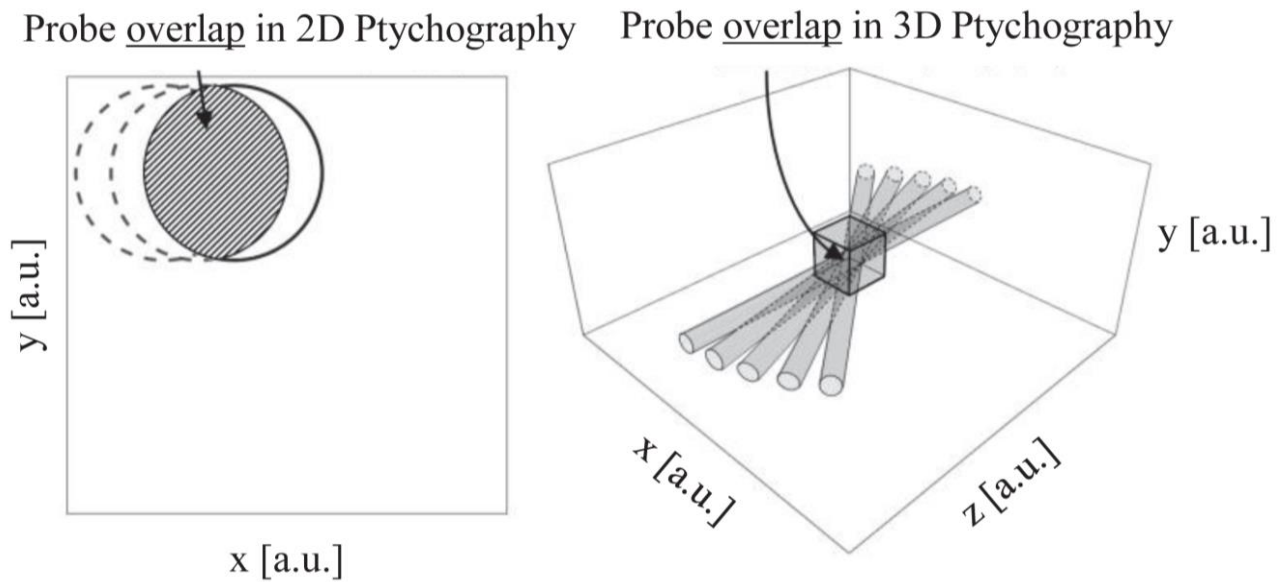

Figure 1: Schematic representation of the overlap between consecutive illuminations in a 2D (left) and 3D (right) ptychography setup. In 2D ptychography, the illumination overlap (represented by the hatched area) is enforced at each tomographic angle by taking scanning step sizes smaller than the probe function width. In a three-dimensional setup, the overlap constraint is applied on illuminations from adjacent projection angles that overlap in the 3D volume that defines the tomographic reconstruction field-of-view.

Recently, Gürsoy [14] has shown that the probe overlap constraint can be significantly relaxed if a combined ptycho-tomography reconstruction approach is applied. In his work, Gürsoy extended the error-reduction algorithm of Fienup [15], [16] to a tomographic setting by intercalating this iterative method with one iteration of the SIRT algorithm [17] for tomographic reconstruction implemented with an additional total-variation (TV) regularization in order to preserve the edges of the object. Besides his proof-of-concept and encouraging results, the method proposed by Gürsoy implicitly relies on the combination of two different optimization strategies with different cost-functions (the error-reduction and the SIRT algorithm). We suggest that improved reconstruction results may be possible if this problem is expressed as a single optimization problem instead of two separate ones with different cost functions. Other independent and relevant works combining phase-retrieval and tomographic reconstruction were developed by Maretzke [18], [19] and by Kostenko [20] applied to full-field imaging methods such as in-line near-field holography with encouraging results demonstrated for real datasets.

In our work, we present a general description of a ptychography experiment and a direct ptycho-tomography reconstruction algorithm based on a single optimization cost-function with the potential to deal with datasets exhibiting *moderate* phase-wrapping. Our reconstruction strategy is based on a non-linear optimization algorithm that we show is able to accurately recover three-dimensional reconstructions of both the real and imaginary part of the sample's complex refractive index, simultaneously.

This document is structured as follows: First, we describe the general forward model for a ptychography experiment. Afterwards, we present our reconstruction technique as a regularized optimization algorithm that solves a quadratic approximation of the Poisson log-likelihood function between the measured intensities and those resulting from our forward model. The behaviour of the proposed algorithm is demonstrated for simulated and real datasets.

At this stage, the proposed reconstruction method relies on a good estimation of the relative alignment parameters between the sample and the detector at each scanning coordinate. In the presence of misalignments, if a reasonable tomographic reconstruction guess is available from the proposed reconstruction method, additional alignment refinement algorithms such as those in [5], [7], [9], [21], [22] could be applied.

## Methods

In a far-field coherent diffraction imaging measurement, the recorded intensity $I_\Theta$ of the diffracted X-ray wavefront at the detector plane can be modelled, in the absence of noise, according to the Fraunhofer approximation:

$$I_\Theta \cong |\mathcal{F}\{\psi_\Theta\}|^2 = |\Psi_\Theta|^2. \qquad (1)$$

Here, $\Theta$ generalizes a set of relative orientation parameters between the incoming wavefront and the sample. These include spatial translations between the illumination and the sample, tomographic rotation angles and other possible angular tilts that may occur during data acquisition. For a monochromatic coherent illumination, the free-space propagation of the exit-wave $\psi_\Theta$, after beam-sample interaction, can be modelled by a simple two-dimensional Fourier transform $\mathcal{F}$. Please note that this approximation is only valid for far-field measurements whereas in the near-field a Fresnel propagator should be used instead.

In our work, we assume that the incoming illumination wavefront, described by a complex-valued probe function, is constant during data acquisition and thus, also for notation simplicity, it will be represented by a unique symbol, $P$. In practice this assumption is not always entirely true and it has been shown that ptychography reconstructions can be improved by refinement algorithms [23] that take local probe variations into consideration. Under the 'thin sample' condition [24], the interaction between the incoming wavefront and the sample can be approximated by a simple multiplicative relation so that

$$\psi_\Theta = PO_\Theta, \qquad (2)$$

where $O$ is the so-called object function and is related to the complex transmissivity of the sample up to a constant phase term. Other first-order phase terms may appear during ptychography reconstructions but are associated with uncertainties in the exact position of the centre of the diffraction patterns [8], [24]. Most of the already existing ptychography algorithms aim to recover both the probe $P$ and the object function $O_\Theta$ in an alternating optimization approach, by iteratively refining the initial guesses of these functions using different, but analogous, strategies [24]–[28]. In turn, $O$ can also be defined as a function of the sample's complex refractive index $n$ as

$$O(x,y) = \exp\left[\frac{2\pi \mathbf{i}}{\lambda}\int(n(x,y,z)-1)\mathrm{d}z\right], \qquad (3)$$

where $\lambda$ represents the wavelength of the incoming X-ray beam, $\mathbf{i}$ is the imaginary unit, $x, y$ and $z$ represent the spatial coordinates of the object expressed in a fixed coordinate system where $z$ is the direction of the incoming (and propagated) parallel X-ray beam, assumed to be perpendicular to the detector plane. For a given refractive index, $n = 1 - \delta + \mathbf{i}\beta$, the local electron density of the sample $\rho_e$ and linear absorption coefficient $\mu$ may be deduced according to

$$\rho_e(x,y,z) = \frac{2\pi\delta(x,y,z)}{r_0\lambda^2}, \qquad (4)$$

$$\mu(x,y,z) = \frac{4\pi\beta(x,y,z)}{\lambda}, \qquad (5)$$

where $r_0 = 2.82 \cdot 10^{-5}$ Å is the Thomson scattering length or classical electron radius.

**Forward model**

The description of our forward model follows naturally from the combination of (1), (2) and (3)

$$I_\Theta^f = F(n') = |\mathcal{F}\{P \cdot \exp[\mathbf{i}k\mathcal{R}_\Theta(n')]\}|^2, \qquad (6)$$

where the wave number is defined as $k = \frac{2\pi}{\lambda}$ and $\mathcal{R}_\Theta$ represents a linear operator that computes the line integral in equation (3) expressed in terms of the relative sample-beam orientation parameters $\Theta$. For notation simplicity we introduced $n' = n - 1 = -\delta + \mathbf{i}\beta$, that is the argument of our forward model and that we aim to resolve by our proposed method.

**Reconstruction algorithm**

As the phase information of the wavefront is lost in data acquisition, the inverse of our forward model $F^{-1}$ cannot be defined uniquely. Also, the presence of noise in the measurements and small differences between the numerical model in equation (6) and the real physical phenomenon, make this inverse problem ill-posed.

The fundamental idea behind our reconstruction algorithm is analogous to those in traditional ptychography methods, meaning that it searches iteratively for a complex solution $\hat{n}'$, for the refractive indices of the sample, that minimize the differences between the experimentally measured intensities $I_\Theta^m$ and those resulting from the forward model in equation (6) $I_\Theta^f$. We choose as our object function the quadratic approximation of the log-likelihood function [29] for the family of normal distributions $N(\mu_N, \sigma_N^2)$ with unknown mean $\mu_N$ and variance $\sigma_N^2$ defined as

$$l(\mu_N, \sigma_N^2, x) = -\frac{1}{2\sigma_N^2}\sum_i(x_i - \mu_N)^2, \qquad (7)$$

where $x_i$ denotes the intensity measurement/observation at the $i$-th pixel. In an X-ray diffraction experiment, the measured intensities follow Poisson statistics, meaning that noise is uncorrelated between pixels and its variance for each pixel can be approximated by the mean measured intensity. Maximizing equation (7) is mathematically equivalent to minimizing the quadratic approximation of the negative log-likelihood, which in turn is expressed as the least-squares problem:

$$\min_{n'} f(n') = \frac{1}{2}\sum_i \left(\frac{I^f_{\Theta_i}(n')-I^m_{\Theta_i}}{\sigma_i}\right)^2. \tag{8}$$

In our implementation the standard deviation in equation (8) is approximated by $\sigma \approx \sqrt{I^m_\Theta + \epsilon}$ in order to decouple any dependence between the noise model and our forward model in (6). Also, the constant $\epsilon = 1$ is used in order to avoid divisions by zero that may introduce numerical errors during reconstruction that are associated with pixels with zero measured intensities. Besides the ill-posed nature of this problem, the forward model (6) is also non-linear because of the squared modulus operator and the exponential function. This suggests that candidate solvers for equation (8) belong to the family of non-linear regularized least-squares minimization algorithms as is the case for the (non-linear, regularized) Gauss-Newton [30], Levenberg-Marquardt [31], [32] and Powell's Dog Leg method, among others. In our work, the problem in equation (8) is solved with the Levenberg-Marquardt algorithm (LMA), also known as damped Gauss-Newton method. Let us define a vector-valued function $r(n') = \frac{I^e_\Theta(n')-I^m_\Theta}{\sqrt{I^m_\Theta+\epsilon}}$, also known as residual vector, so that

$$\frac{1}{2}\sum_i \left(\frac{I^f_{\Theta_i}(n')-I^m_{\Theta_i}}{\sqrt{I^m_\Theta+\epsilon}}\right)^2 = \frac{1}{2}r(n')^*r(n'). \tag{9}$$

Here the superscript '∗' represents the adjoint or conjugate transpose operator. The gradient and Hessian of the cost function $f(n')$ are expressed as

$$\nabla f(n') = J(n')^*r(n'), \tag{10}$$
$$\nabla^2 f(n') = J(n')^*J(n') + Q(n'), \tag{11}$$

where $J(n') = \frac{\partial r}{\partial n'} = \frac{1}{\sqrt{I^m_\Theta+\epsilon}}\frac{\partial I^f_\Theta(n')}{\partial n'}$ is the Jacobian of $r(n')$, and $Q(n')$ denotes higher order quadratic terms of the Hessian that are often ignored for many large scale problems to improve computational efficiency. The cost function $f(n')$ can be linearized around a current estimate of a minimizing point by means of a second-order Taylor expansion expressed as a function of the Jacobian and Hessian as

$$f(n' + \Delta n') \approx f(n') + \nabla f(n')\Delta n' + \frac{1}{2}\Delta n'^* \nabla^2 f(n')\Delta n'. \tag{12}$$

As most numerical minimization algorithms, the LMA adjusts an initial guess for $n'$ by taking an update step $h_{n'} = \Delta n'$ that minimizes the quadratic cost-function. This is done by differentiating (12) and computing its roots. This way, at the $l$-th iteration, we define:

$$h^{[l]}_{n'} = -\left(\nabla^2 f(n'^{[l]}) + \lambda_{lm}\right)^{-1} \nabla f(n'^{[l]}), \tag{13}$$
$$n'^{[l+1]} = n'^{[l]} + h^{[l]}_{n'}. \tag{14}$$

In equation (13), $\lambda_{lm}$ is a damping factor, introduced by Levenberg [31] that regularizes the Hessian and is updated at each iteration. Considering the aforementioned first-order approximation of the Hessian, both $\nabla f(n')$ and $\nabla^2 f(n')$ require the definition of $J(n')$ and its adjoint $J(n')^*$. As in the work by Maretzke [18], [19] we express these functions implicitly, depending on the current reconstruction $n'^{[l]}$:

$$J\left(n'^{[l]}\right)h^{[l]}_{n'} = \frac{1}{\sqrt{I^m_\Theta+\epsilon}}\Re\left[\overline{(\mathcal{F}\{P \cdot \exp[\mathbf{i}k\mathcal{R}_\Theta(n'^{[l]})]\})}\mathcal{F}\left\{\mathbf{i}kP \cdot \exp\left[\mathbf{i}k\mathcal{R}_\Theta\left(n'^{[l]}\right)\right]\mathcal{R}_\Theta\left(h^{[l]}_{n'}\right)\right\}\right] =$$
$$\frac{1}{\sqrt{I^m_\Theta+\epsilon}}\Re\left[\overline{(\Psi^{[l]}_\Theta)}\mathcal{F}\left\{(\mathbf{i}k\psi^{[l]}_\Theta)\mathcal{R}_\Theta\left(h^{[l]}_{n'}\right)\right\}\right], \tag{15}$$

$$J\left(n'^{[l]}\right)^* h_g^{[l]} = \mathcal{R}_\Theta^* \left[\overline{\imath k P \cdot \exp[\imath k \mathcal{R}_\Theta(n'^{[l]})]} \mathcal{F}^{-1} \left\{\frac{1}{\sqrt{I_\Theta^m + \epsilon}} \left(\mathcal{F}\left\{P \cdot \exp\left[\imath k \mathcal{R}_\Theta\left(n'^{[l]}\right)\right]\right\}\right) \Re\left[h_g^{[l]}\right]\right\}\right] =$$
$$2\mathcal{R}_\Theta^* \left[\overline{\left(\imath k \psi_\Theta^{[l]}\right)} \mathcal{F}^{-1} \left\{\frac{\Psi_\Theta^{[l]} \Re[h_g^{[l]}]}{\sqrt{I_\Theta^m + \epsilon}}\right\}\right]. \quad (16)$$

In equations (15) and (16) the overbar denotes complex conjugation, $\Re$ is the real-part (self-adjoint) operator and $\mathcal{R}_\Theta^*$ generalizes the tomographic *back-transform* according to the fields-of-view and sample orientation parameters described by $\Theta$.

**Probe Retrieval**

In most CDI experiments, the probe function $P$ may not be known or available, but it can be retrieved by ptychography algorithms in an alternating optimization approach. In our work we follow the same method for reconstructing $n'$ in order to recover $P$, intercalating both optimization problems after each iteration of the proposed algorithm. The probe reconstruction problem is then formulated as

$$\hat{P} = \underset{P}{\operatorname{argmin}} \frac{1}{2} \sum_i \left(\frac{I_{\Theta_i}^f(P) - I_{\Theta_i}^m}{\sigma_i}\right)^2. \quad (17)$$

Here, the Jacobian and Hessian approximations are expressed as functions of $J_p(P)$ and $J_p(P)^*$, that in turn are given by

$$J_P(P^{[l]}) h_P^{[l]} = \frac{1}{\sqrt{I_\Theta^m + \epsilon}} \Re\left[\overline{\left(\Psi_\Theta^{[l]}\right)} \mathcal{F}\left\{O_\Theta^{[l]} h_P\right\}\right], \quad (18)$$

$$J_P(P^{[l]})^* h_g^{[l]} = 2\overline{O_\Theta^{[l]}} \mathcal{F}^{-1} \left\{\frac{\Psi_\Theta^{[l]} \Re[h_g^{[l]}]}{\sqrt{I_\Theta^m + \epsilon}}\right\}. \quad (19)$$

**Discretization and Implementation**

The linear operator $\mathcal{R}_\Theta$ and its adjoint were implemented using the ASTRA toolbox [33] in order to exploit GPU resources for faster computations on large datasets. Besides, a full description of the relative positioning between incoming beam, sample and detector is allowed, enabling the reconstruction of datasets acquired with more complex scanning geometries. This *flexible* operator can also be used to correct for any known translational or angular motion that the sample may experience during measurements *i.e.* vibrations or wobbling of the rotation axis. The 2D Fourier transform $\mathcal{F}$ and its adjoint/inverse were implemented using the FFTW library [34] and all multiplications in equations (6), (14) and (15) are to be understood element-wise. The developed algorithms were implemented and tested using MATLAB®2017a and Python2 on both Windows and Linux platforms and are publically available online under the DOI: 10.6084/m9.figshare.6608726.

Each iteration of the LMA, computes the solution update step in equation (13) by solving the least-squares problem:

$$\left(J^{*[l]} J^{[l]} + \lambda_{lm} \mathbf{I}\right) h_{n'}^{[l]} = -J^{*[l]} r^{[l]}, \quad (20)$$

using the Hestenes-Stiefel version of the conjugate gradient method (CGM)[35], [36]. In equation (20), $\mathbf{I}$ is the identity matrix. In turn, the damping term of the LMA, $\lambda_{lm}$ is updated at each iteration using the strategy in [37].

The application of constraints in $n'$, after each solution update, is seen to be essential for accurate quantitative reconstruction of the refractive indices of the sample. These become particularly important in order to resolve datasets with wrapped phases in the object function (3). In our implementation, a non-negativity constraint is applied to $n'$, after each iteration, limiting the domain of $\delta$ and $\beta$ to only non-negative numbers. Once a new solution for $n'$ is computed we write

$$n'^{[l]} = -\delta^{[l]} + \mathbf{i}\beta^{[l]} = \min(-\delta^{[l]}, 0) + \mathbf{i}\max(\beta^{[l]}, 0). \quad (21)$$

The resulting reconstruction algorithm is summarized in the pseudocode Algorithm 1:

---

**Algorithm 1: Tomographic Reconstruction of Far-Field Coherent Diffraction Patterns**

---

1. *Initialize* $n'$, $\lambda_{lm}$
2. **While** $l < l_{max}$ or not stop-criterion **do**:

   *Use forward model to compute*
   $F^{[l]} = F(n'^{[l]})$

   **Solve** *(12) using CGM (Hestenes-Stiefel):*
   $(J^{*[l]}J^{[l]} + \lambda_{lm}\mathbf{I})h_{n'}^{[l]} = -J^{*[l]}F^{[l]}$

   **Update** *sample solution*
   $n'^{[l+1]} = n'^{[l]} + h_{n'}^{[l]}$

   **Enforce constraints** *to sample solution*

   **Update** $\lambda_{lm}$

   **Solve** *using CGM (Hestenes-Stiefel):*
   $\left(J_P^{*[l]}J_P^{[l]} + \lambda_{lm,P}\mathbf{I}\right)h_P^{[l]} = -J_P^{*[l]}F^{[l]}$

   **Update** *probe solution*
   $P^{[l+1]} = P^{[l]} + h_P^{[l]}$

   **Enforce constraints** *to Probe solution*

   **Update** $\lambda_{lm,P}$

   **End While**

---

## Simulation setup

A $300^3$-voxel discretized complex-valued phantom was used to generate 4000 diffraction patterns randomly distributed over 180 degrees around the rotation axis of the sample. These were computed using the forward model in equation (6) and afterwards rounded to integer numbers. Four different simulations were conducted in order to evaluate the robustness of our algorithm in the presence of noise and *weakly-absorbing* samples. In this context, *weakly-absorbing* samples are understood as those with $\beta \ll \delta$. The simulated $\delta$ and $\beta$ values are in the range of 0 to $10^{-5}$ for two of the simulations, whereas for the *weakly-absorbing* samples the $\beta$ ranges from 0 to $10^{-7}$. The detector with $100^2$ pixels of 172 μm was placed at 5 m from the sample, which for a wavelength of 1 Å corresponds to a reconstructed voxel size of 29 nm. The probe function was generated using a Gaussian function, with a total integrated intensity of $10^7$ photons and is illustrated in Figure 2. The generated $\delta$ and $\beta$ volumes are depicted in Figure 3 for the *weakly-absorbing* sample case. In this work the real and imaginary part of the simulated phantom were generated independently with different structures. This was done in order to demonstrate a good reconstruction performance for cases where no explicit dependence between $\delta$ and $\beta$ may be expressed.

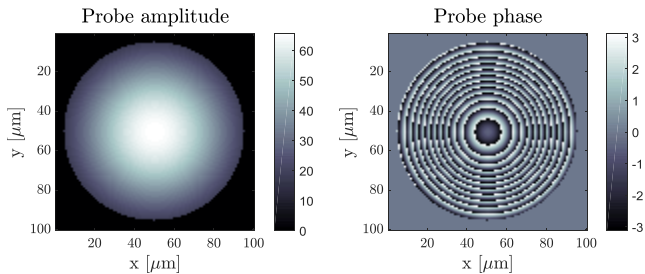

**Figure 2: Amplitude and phase of the complex-valued probe function used in our simulations.**

When imaging samples with high $\delta$ and/or large thicknesses, phase shifts of the object function (equation (3)) may be larger than $\pi$ resulting in wrapping of the phase. Current reconstruction strategies rely either on phase-unwrapping algorithms [38]–[40] of the phase-contrast projections prior to tomographic reconstruction, or alternative tomographic reconstruction algorithms such as those in [8], [9]. In order to evaluate the robustness of our reconstruction method in the presence of phase-wrapping a simulated sample with $\delta$ values in the range of $[0 \sim 5 \times 10^{-5}]$ was used resulting in phase-wrapping of the object function as shown in Figure 5b.

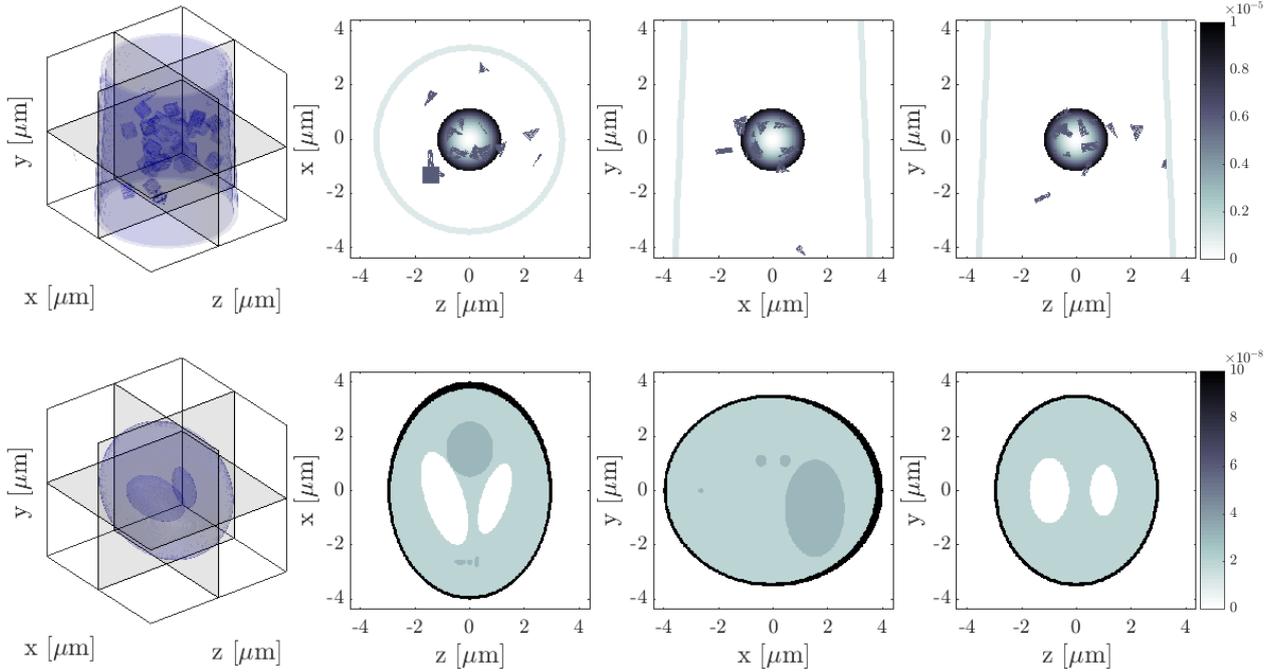

Figure 3: Complex-valued phantom used in our simulations. On the top: 3D isosurface and cross-section over cardinal planes for simulated $\delta$ object. On the bottom: Same for simulated $\beta$ volume.

All tomographic reconstructions presented in this document were obtained using 25 iterations of the CGM and 50 iterations of the LMA, except for the datasets with phase-wrapping where 100 iterations of the LMA were used instead.

## Results and Discussion

The final reconstructions of simulated data, using our proposed reconstruction algorithm, are presented in Figure 4. Our results suggest that Poisson noise in the measurements decreases the reconstruction quality as is intuitively expected, and is responsible for artefacts during the initial reconstruction iterations. Such artefacts arise from erroneous contributions of the real part of the solution update $h_N$ into its imaginary part or vice versa. We also found that reconstructing $\beta$ for weakly-absorbing samples is increasingly challenging as the results in Figure 4 illustrate. We believe that this effect can be explained by the difference in orders of magnitude between the $\delta$ and $\beta$ values of the sample. When such difference is *large* our reconstruction method is mostly dominated by updates of the $\delta$ volume that have a higher impact on the cost-function in equation (8). On the other hand, when the $\beta$ and $\delta$ are in the same order of magnitude, the reconstructed tomograms are in good agreement with both the real and imaginary parts of the sample's refractive indices.

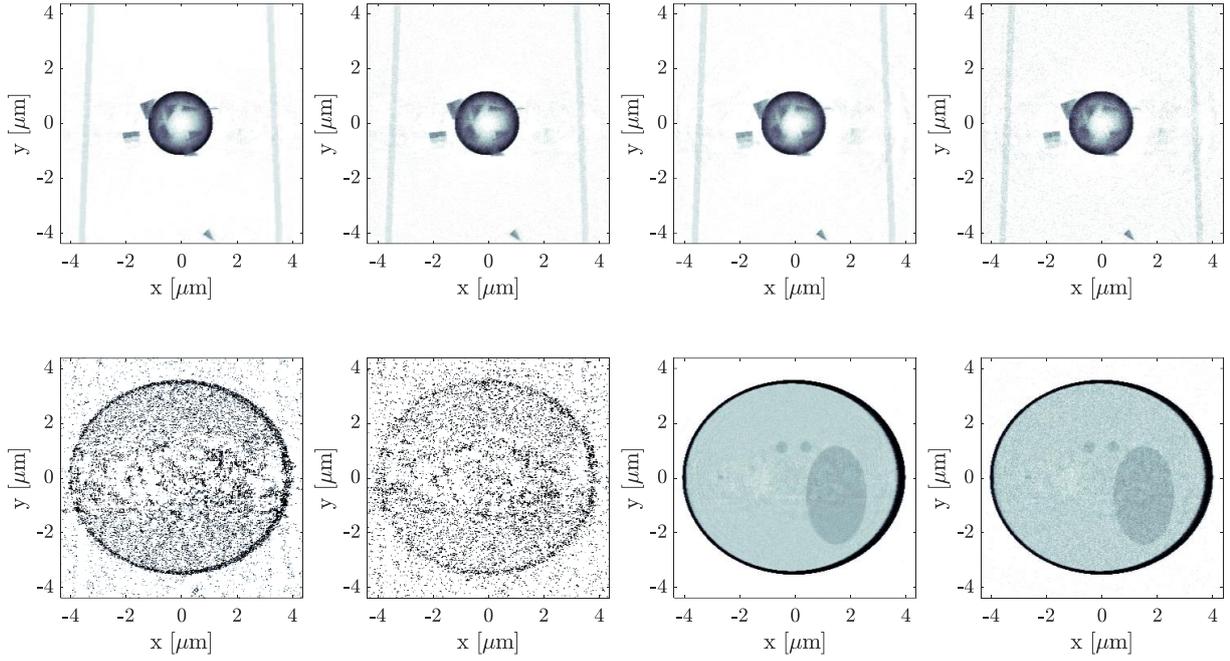

**Figure 4: X-Y tomographic slices of the reconstructed refractive index for simulation data. The reconstructed $\delta$ volumes are represented at the top and the $\beta$ on the bottom. From left to right: low-absorption noise-free; low-absorption Poisson noise; high-absorption noise-free; high-absorption Poisson noise. The intensity scale in each image was properly normalized according to the refractive index of the sample i.e. $[0 \sim 10^{-5}]$ for all images except for the $\beta$ representations with low-absorption where $[0 \sim 10^{-7}]$ was used instead.**

The full proof of convergence of our proposed method in the presence of large phase-shifts ($> \pi$) is not part of the scope of this publication. Nonetheless, our simulations indicate that by the use of appropriate constraints, such as the one in equation (21) and at the cost of additional iterations, accurate tomographic reconstructions can be achieved by the reconstruction algorithm as exhibited in Figure 5.

An example of application of the proposed algorithm to a real dataset is presented in Figure 6. The diffraction data was acquired at the cSAXS beamline from the Paul Scherrer Institut from a LiFePO$_4$ powder sample inside a glass capillary tube. A total of 200 ptychography projection images were reconstructed by the difference map algorithm with maximum-likelihood probe positioning refinement [23], [27], [41] from 172 different scanning positions each. As our proposed reconstruction algorithm requires handling all diffraction patterns at once, we saw ourselves limited by its high memory demands both in terms of RAM and GPU onboard memory. For this reason, we show the behaviour of our reconstruction method by using only a small fraction of the total acquired dataset. For the reconstruction in Figure 6c the diffraction patterns were cropped to their central 300 × 300 pixels and in Figure 6d to their central 400 × 400 pixels. Please note that this operation results in an increase of the reconstructed voxel size from 14.3 nm to 28.6 nm and 21.5 nm for Figure 6c and Figure 6d respectively, and consequently to a decrease in spatial resolution. Furthermore, the reconstruction in Figure 6d was obtained by using only 1/4$^{th}$ of the total diffraction patterns. Due to computational limitations, our reconstruction algorithm was run with only 20 iterations for both CGM and LMA algorithms, which in this case were sufficient to illustrate the main features of the sample inner structure. Improvements in reconstruction accuracy and resolution could be achieved, for example, by increasing the number of iterations taken in both CGM and LMA, and the amount or size of the diffraction data used. We expect that it will soon be possible to handle full data sets with the advent of memory architectures linked across several GPU's.

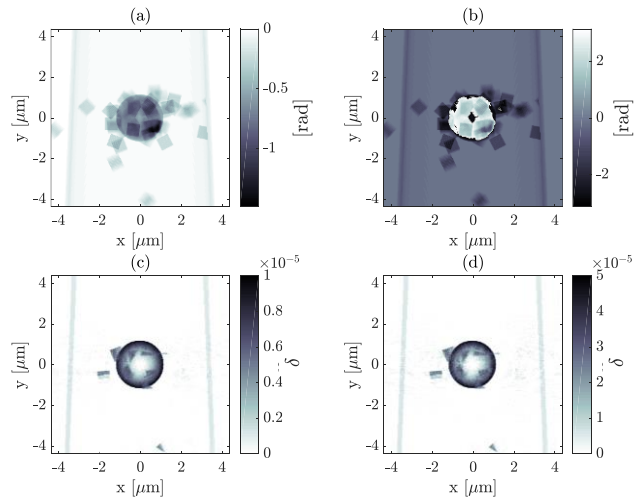

**Figure 5: Comparison between tomographic reconstructions, by the proposed algorithm, from datasets with and without phase-wrapping of the object function. (a) and (b) illustrate the phase-shifts of the object function without and with phase-phase-wrapping respectively. (c) and (d) show X-Y slices from tomographic reconstructions exhibiting an accurate reconstruction of $\delta$ for the dataset containing wrapped phases (or large phase-shifts).**

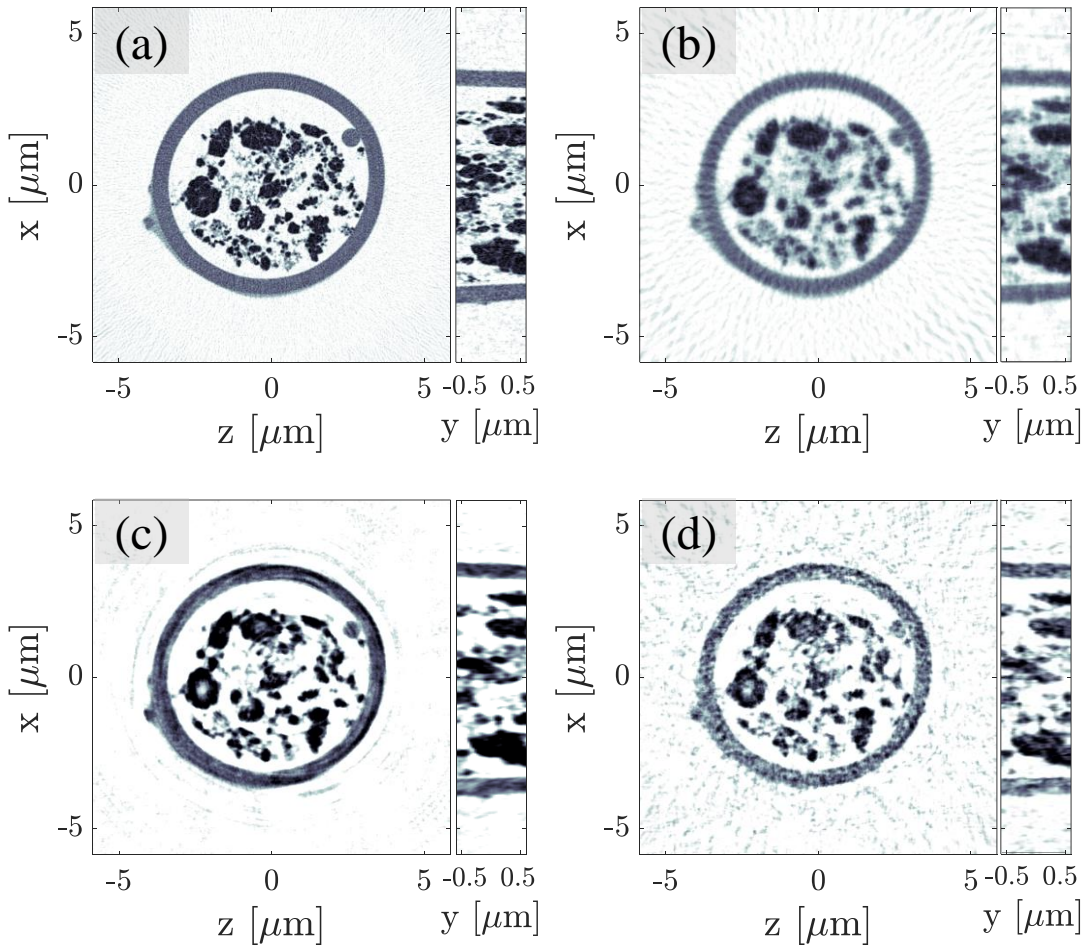

Figure 6: Tomographic reconstructions of $\delta$ values obtained by different reconstruction approaches: (a) Filtered backprojection of 200 uniformly distributed projections. (b) Tomographic reconstruction with 1/4$^{th}$ of the dataset (50 projections). Because of the reduced number of projections the reconstruction was obtained by the SIRT algorithm with 200 iterations. (c) Tomographic reconstruction by the proposed algorithm with 20 iterations of the CGM and 15 iterations of the LMA algorithm. (d) Tomographic reconstruction by the proposed algorithm with 20 iterations of the CGM and 20 iterations of the LMA algorithm with 1/4$^{th}$ of the total diffraction patterns dataset. Reconstructed voxel sizes of 14.3 nm for for (a) and (b), 28.6 nm for (c) and 21.5 nm for (d). The greyscale varies linearly with the $\delta$ values from 0 (white) to $1 \times 10^{-5}$ (black).

# Conclusions and future work

In this work we have presented a numerical algorithm for direct tomographic reconstruction of the volumetric distribution of the refractive index of a sample from intensity measurements of far-field coherent diffraction patterns. This non-linear and ill-posed inverse problem is framed as a least-squares optimization problem, taking into account Poisson noise statistics in the measured intensities, which we solve by the Levenberg-Marquardt algorithm. Our simulation results show that accurate reconstruction of the sample's refractive indices can be retrieved by the proposed method, which is of special interest for phase-contrast tomography experiments. Our studies also indicate that the convergence of this proposed reconstruction method is mostly dominated by the highest values of $n'$, making the reconstruction of the absorption properties of the sample increasingly challenging for the cases where $\beta \ll \delta$.

Future improvements to our proposed algorithm will include the implementation of other regularization methods, such as Tikhonov (for noise suppression) and total variation (for edge-enhancement) in our optimization problem. Additional constraints in both direct and/or reciprocal space may also benefit the convergence of the proposed reconstruction method by restricting/bounding the solution space of our problem. We are currently working on a multiscale reconstruction approach to decrease the total computational time by first reconstructing a low-resolution version of the object that is then taken as initial solution guess for the high-resolution reconstruction.

So far, all the presented tomographic reconstructions relied on the exact knowledge of the probe function and the sample spatial coordinates at each scanning position. The extension of the proposed method for probe retrieval is an ongoing endeavour.